\RequirePackage{fix-cm}
\documentclass[smallextended]{svjour3}       
\smartqed  
\usepackage{graphicx}

 \usepackage{amsmath,bm}
 \usepackage{color}
\definecolor{g-blue}{rgb}{0.83,0.95,1}
\definecolor{g-yellow}{rgb}{1,1,0.7}
\definecolor{g-green}{rgb}{0.9,1,0.9}
\definecolor{green}{rgb}{0,0.6,0}
\definecolor{cyan}{rgb}{0,0.7,0.7}
\definecolor{black}{rgb}{0,0,0}
\definecolor{grey}{rgb}{0.4 ,0.4 ,0.4 }

\newcommand{\B}[1]{{\bm{#1}}}
\newcommand{\C}[1]{{\mathcal{#1}}}    

\renewcommand{\sb}[1]{_{\text {#1}}}  
\newcommand{\Eq}[1]{Eq.\,(\ref{#1})}

\def\He4 {$^4$He~}

\begin{document}

\title{Coupled dynamics for superfluid \He4 in the channel.
}


\author{D. Khomenko \and P.  Mishra  \and A. Pomyalov
}


\institute{D.  Khomenko  \and P. Mishra \and A. Pomyalov  \at
              Department of Chemical Physics, Weizmann Institute of Science, 76100 Rehovot, Israel\\
              \email{dmytro.khomenko@weizmann.ac.il}           
}

\date{Received: date / Accepted: date}

\maketitle

\begin{abstract}

 We study the coupled dynamics of normal and superfluid components of the superfluid $^4$He in the channel considering   the counterflow turbulence with laminar normal component.  In particular, we  calculated profiles of the normal velocity, the mutual friction, the vortex line density and  other flow properties  and compared them to the case when the dynamic of the normal component is "frozen".  We have found that the coupling between the normal and superfluid components  leads to flattening of the normal velocity profile, increasingly more pronounced with temperature, as the mutual friction, and therefore coupling, becomes stronger. The commonly measured flow properties also change when the coupling between two components is taken into account.

\end{abstract}

\section{Introduction}
\label{intro}

Superfluid $^4$He below transition temperature $T_\lambda\simeq 2.17\,$K may be viewed as a two-fluid system\cite{PNAS_intro,VinenNiemela,SS-2012} consisting of  normal fluid with the very low kinematic viscosity $\nu\sb n(T)$ and inviscid superfluid component, that have their own densities, $\rho\sb n(T)$, $\rho\sb s(T)$, and velocity fields, $\B u\sb n(\B r,t)$, $\B u \sb s(\B r,t)$.
Due to quantum mechanical restriction, the circulation around the superfluid vortices is equal to $\kappa= h /m\simeq 10^{-3}\,$cm$^2$/s, where $h$ is the Plank constant and $m$ denotes the mass of $^4$He atom.
The singly quantized vortices  usually arrange themselves in a tangle, referred to as  \emph{quantum turbulence}, that may be characterized by vortex line density (VLD) $\cal L$, i.e., total length of the quantized vortex line in a unit volume.

The large scale motion of such a system may be described by Hall-Vinen-Bekarevich-Khalatnikov equations(HVBK)\cite{HV,BK}, where both components are considered as continuous fluids, coupled by a mutual friction force. The microscopic description of the vortex lines dynamics at scales that are still much larger than the vortex core size $a_0\approx 10^{-8}$ cm,  was proposed by Schwarz\cite{Schwarz85,Schwarz88} in the framework of Vortex Filament Method (VFM). In his approach, the influence of the normal component the dynamics on the quantum vortex lines is accounted through the mutual friction force with the predefined time-independent normal fluid velocity. It was soon realized that the back-reaction of the  vortex tangle on the dynamics of the normal component  may be important and several self-consistent methods, coupling Navier-Stokes equation for the normal component with  Lagrangian description of the vortex lines dynamics, were
proposed\cite{BarSam99,Idowu00,KBS_science00}. These methods are computationally challenging, given the wide range of scales involved in the problem. Such a self-consistent studies were limited to space-homogeneous flows with diluted vortex tangles.

The VFM methods were extended to the wall-bounded flows\cite{Samuels92,AaW94} and more recently included  full Biot-Savart description of the tangle \cite{BL13,BL15,YT15,Dima15}. Also here, a time-independent mean normal velocity profile, or a snapshot of the turbulent velocity field were imposed to generate the quantum vortex tangle.

Depending on the way the quantum turbulence is generated in the channel or pipe, the normal and superfluid  components flow in the same direction, or in  opposite directions. In the latter case (the thermal counterflow),
a relative, counterflow velocity $V\sb{ns}$ is established in the channel.
At sufficiently large values of  $V\sb{ns}$ the quantum vortex tangle is created.  As  $V\sb{ns}$ increases, the thermal counterflow passes several stages. At relatively low values of counterflow velocity, the normal components remains laminar, while a fully-developed quantum vortex tangle is formed, indicating superfluid turbulence. This regime was labeled at T1 state \cite{Tough_book}. At higher values of $V\sb{ns}$, the normal component also becomes turbulent, forming so-called T2 state.

In this paper we address the back-reaction of the quantum vortex tangle on the normal velocity profile,
restricting our study to T1 state of the counterflow in the channel. Here we describe the laminar normal velocity by its mean profile (neglecting normal velocity fluctuations), but allow its evolution, driven by the mutual friction force that couples dynamics of the mean normal velocity with the evolution of the vortex tangle. To be consistent with the nature of the mean profile, the mutual friction force, generated by the vortex tangle, is integrated over space and short intervals of time. Such a multi-scale, multi-time approach allows us to follow the coupled dynamics of two components and account for the influence of the vortex tangle on the normal component.

We have found that  initially parabolic mean normal velocity profile evolves to a flatter shape, with effect stronger for higher temperature and counterflow velocities. As a consequence, the VLD profile becomes more uniform in the channel core with the peaks pushed towards the wall. Both the global and microscopic  properties of the flow change compared to the uncoupled dynamics, although this effect is  significant only at high temperatures.
\begin{table}[b]
\caption{Parameters of Heluim-II,used in simulations\cite{ExpData}. }
\label{tab:1}       
\begin{tabular}{|c|c|c|c|c|c|}
\hline\noalign{\smallskip}
T(K) & $\alpha$ & $\alpha'$ & $\rho_n$  (g/cm$^3$) & $\rho_n/\rho_s$ & $\nu=\eta/\rho_n$ (cm$^2$/sec) \\
\noalign{\smallskip}\hline\noalign{\smallskip}
1.45 & $6.1 \times 10^{-2}$ & $1.8 \times 10^{-2}$ & $1.3 \times 10^{-2}$ & $1.0 \times 10^{-1}$ & $1.0 \times 10^{-3}$\\
1.6 & $9.7 \times 10^{-2}$ & $1.6 \times 10^{-2}$ & $2.4 \times 10^{-2}$ & $1.9 \times 10^{-1}$ & $5.6 \times 10^{-4}$\\
1.9 & $2.1 \times 10^{-1}$ & $8.3 \times 10^{-2}$ & $6.1 \times 10^{-2}$ & $7.2 \times 10^{-1}$ & $2.2 \times 10^{-4}$\\
\noalign{\smallskip}\hline
\end{tabular}\label{TabPar}
\end{table}
\section{Coupled system}
\label{sec:1}
 In the framework of two-fluid description of the superfluid $^4$He, dynamics of the normal component is given by HVBK equation\cite{HV,BK,BarSam99}:
\begin{equation}\label{eq:NS}
\frac{\partial \bm u\sb n}{\partial t}+ (\bm u\sb n \cdot \nabla)\bm u\sb n = \frac{\nabla P}{\rho\sb n} + \frac{\bm F\sb {ns}}{\rho\sb n} +\nu \sb n \Delta \bm u\sb n,
\end{equation}
where mutual friction force $\bm F\sb {ns}$ couples the two components and the effective pressure $P$ is defined by: $\nabla P=-\rho\sb n/\rho  ~\nabla p +\rho\sb s S \nabla T$  ($S$ is entropy and $\rho=\rho\sb n+\rho\sb s$ is density of Helium II).
For the laminar flow in the channel, \Eq{eq:NS} simplifies to an equation for the mean normal velocity:
\begin{equation}\label{VnEq}
\frac{\partial V\sb n(y)}{\partial t}= \frac{d P}{d x} + \frac{\C F\sb {ns}(y)  }{\rho\sb n}+\nu\sb n \frac{\partial^2 V_n(y)}{\partial y^2}  ,
\end{equation}
Here we took into account  that in the planar channel geometry the normal velocity field $V\sb n(y)$ depends  on the  wall-normal direction only and has only one non-zero (streamwise) component along channel. The mutual friction term $\C F\sb {ns}(y)$ is a time average of a streamwise projection of the  mutual friction force,
\begin{equation}\label{eq:Fns}
\bm F\sb{ns}= \frac{\rho_s }{\Omega} \int_{\C C'} (\alpha \bm s' \times [\bm s' \times \bm V\sb{ns}]+\alpha' \bm s' \times \bm V\sb{ns} )d \xi,
\end{equation} defined by dynamics of the quantum vortex tangle\cite{Schwarz88}.
Here $\bm s$ is a radius vector to  the points on the vortex line, $'$ denotes derivative along the vortex line and $\alpha, \alpha'$ are the mutual friction coefficients. The integral in \Eq{eq:Fns} is taken along vortex lines $\C C'$, residing inside suitably defined  coarse-grained volume $\Omega$.

The instantaneous  counterflow velocity $\bm V\sb{ns}=\bm {V}\sb n-\bm {V}\sb s$ is defined by velocities of the normal and superfluid components. The superfluid velocity $\bm V\sb s =  {V\sb s ^0}+ \bm {V}\sb {BS}(\bm s)$
contains contributions of the  vortex tangle  velocity $\bm {V}\sb {BS}(\bm s)$  in the Biot-Savart representation, integrated over entire vortex configuration $\C C$ and the applied superfluid velocity ${V\sb s ^0}$, defined by the counterflow condition $\rho\sb n \langle \bm {V}\sb n \rangle_v +\rho\sb s \langle \bm {V}\sb s\rangle_v =0$. Here $\langle...\rangle_v$ stands for global volume averaging.
\begin{figure*}[t]
\begin{tabular}{ccc}
\includegraphics[width=0.42\textwidth]{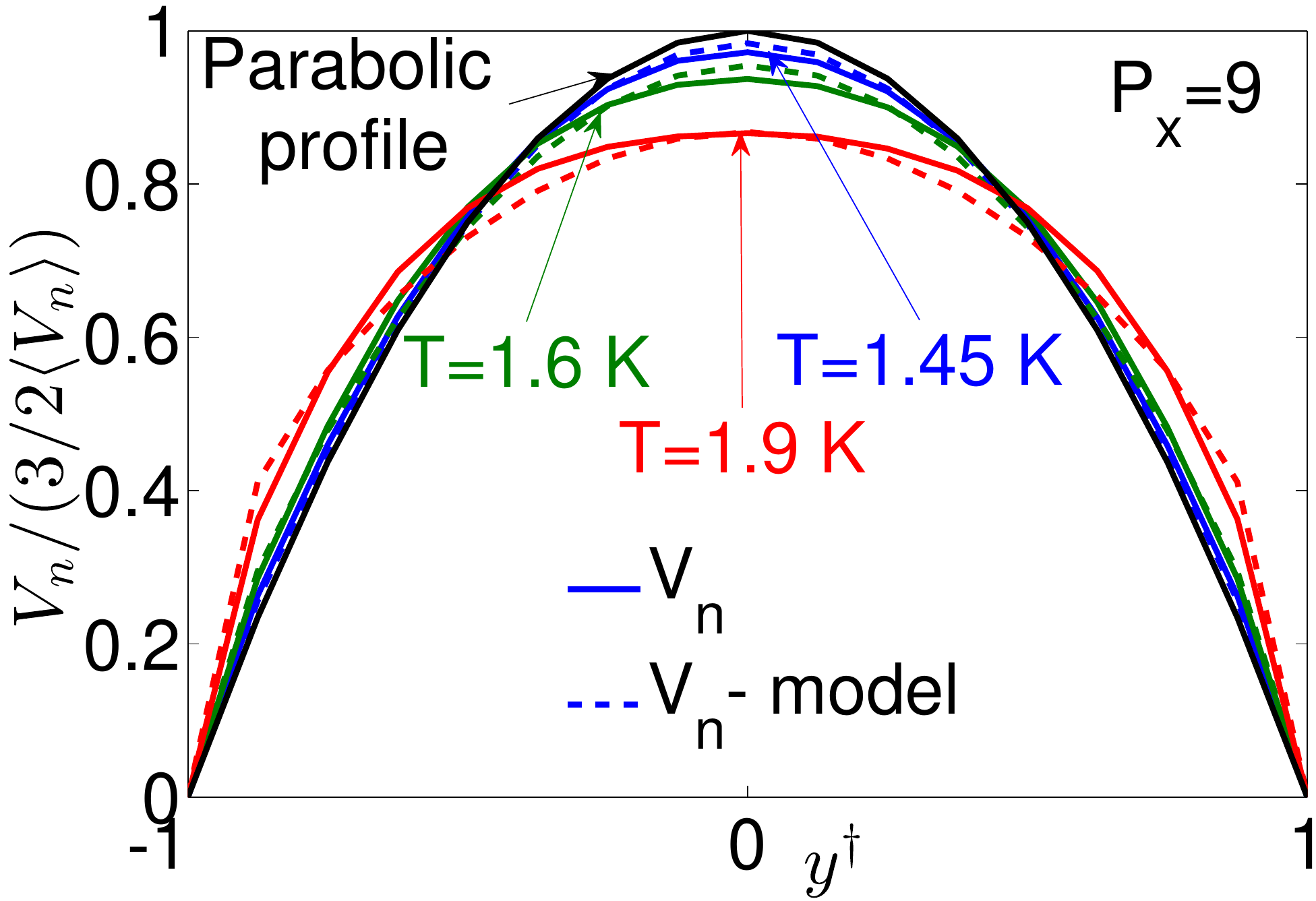}&
\includegraphics[width=0.42\textwidth]{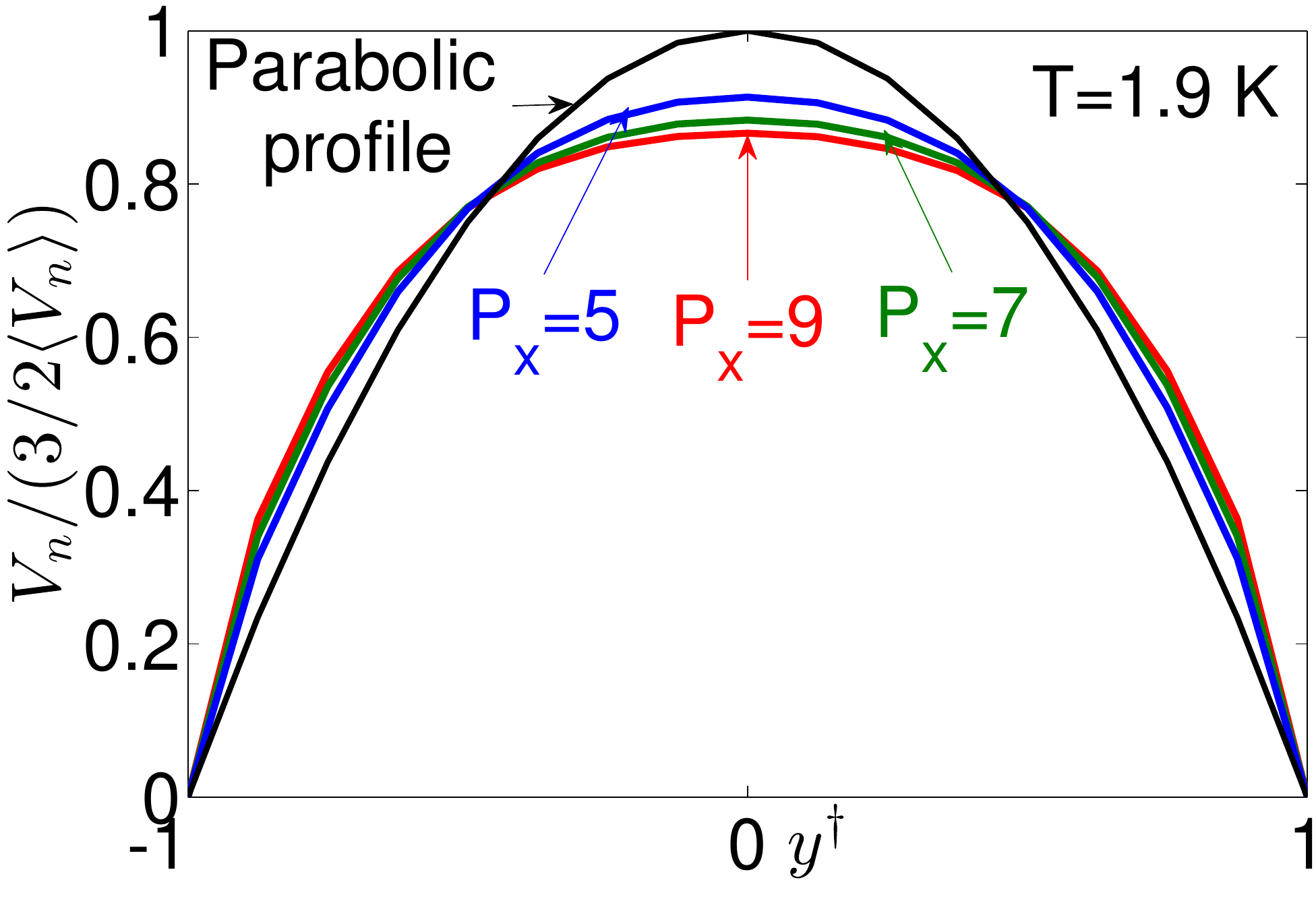}&
\end{tabular}
  \caption{\label{fig:Vnprofiles}Comparison of normalized velocity profiles $V\sb n(y)/(3/2~\langle V\sb n\rangle)$ vs $y^\dagger=y/(h), y^\dagger \in[-1,1]$ for effective pressure gradient $P\sb x=9$ and different temperatures (left panel) and for $T=1.9$K and different values $P\sb x$ (right panel). The normalized parabolic profile is shown by black solid line. In the left panel we also show the model profiles Eq. \eqref{Vn4}(dashed lines) for  $T=1.45$K[$\langle V\sb n\rangle=1.22$ cm/s], $1.6$K[$\langle V\sb n\rangle=1.01$ cm/s]and $1.9$K[$\langle V\sb n\rangle=0.69$cm/s]. Values of $A\sb{GM}$, calculated from mutual friction force in the coupled dynamics are: $A\sb{GM}(1.45)=20.7$cm$\cdot$s/g, $A\sb{GM}(1.6)=35.7$cm$\cdot$s/g, $A\sb{GM}(1.9)=47.2$cm$\cdot$s/g and $n=14$.(Color Online) }
\end{figure*}
\begin{figure*}[b]
\begin{tabular}{cc}
\includegraphics[width=0.42\textwidth]{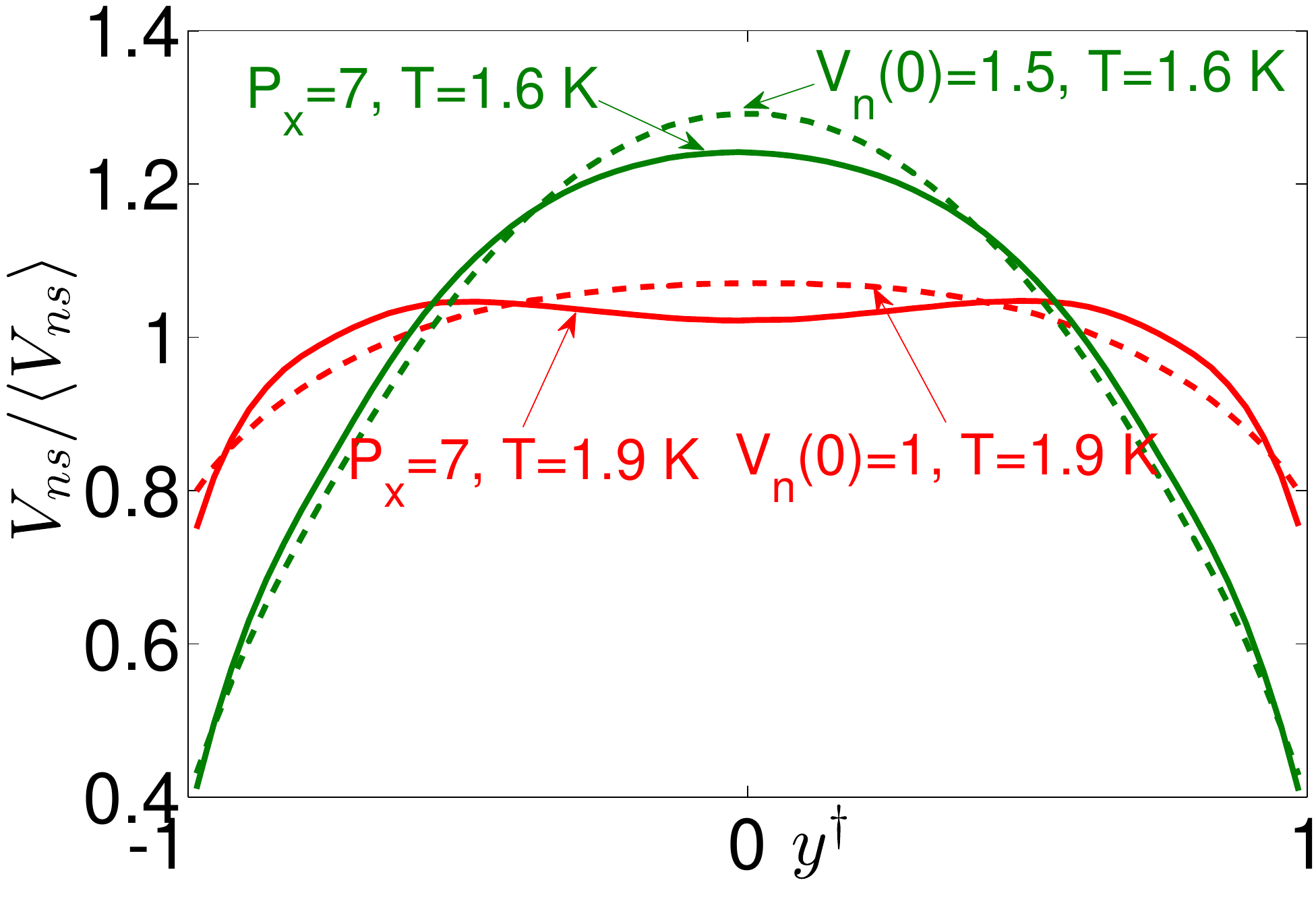}&
\includegraphics[width=0.42\textwidth]{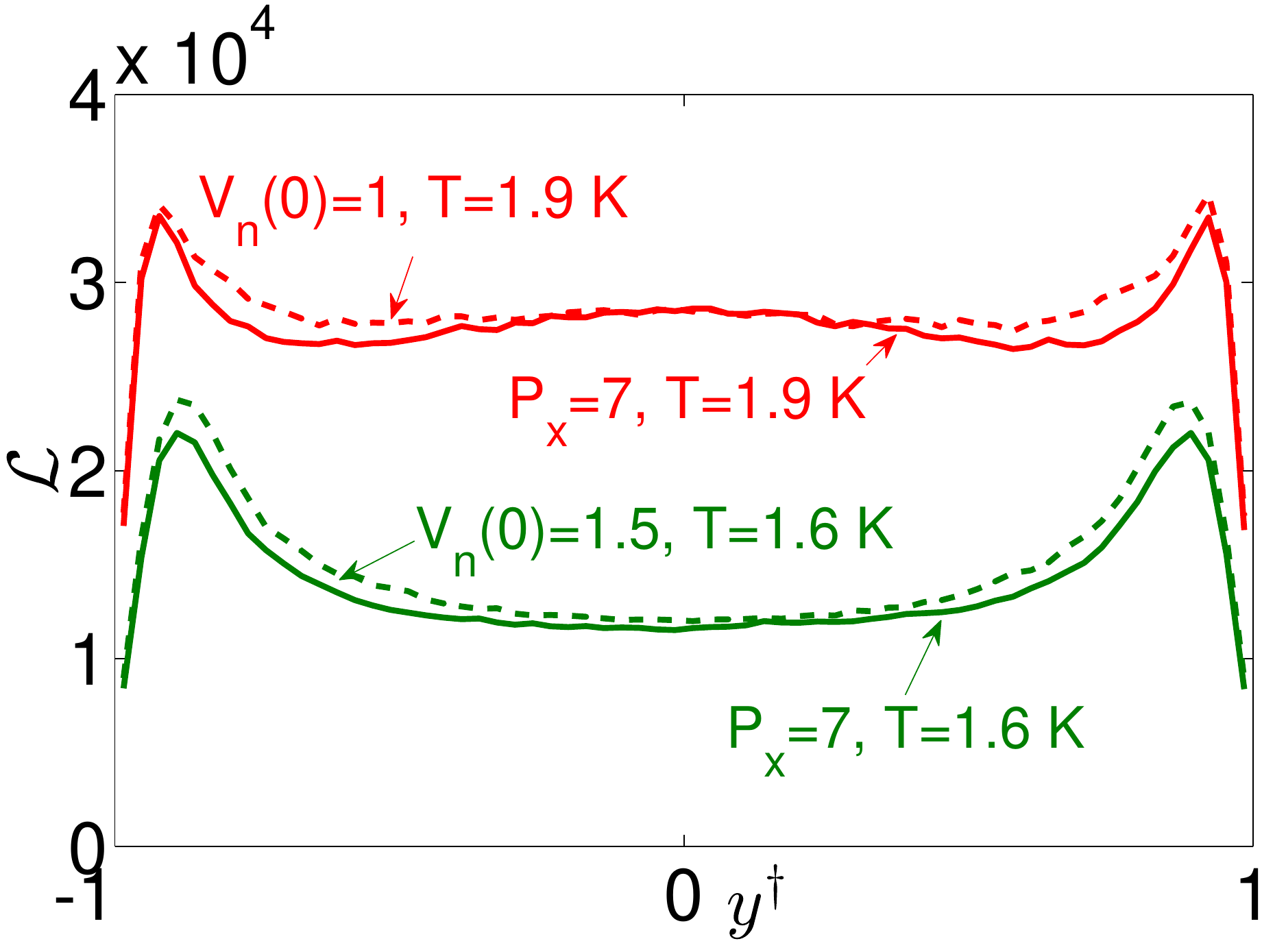}
\end{tabular}
 \caption{\label{fig:Lprofiles} Comparison of normalized $V\sb{ns}$(left) and VLD profiles(right) for coupled dynamics  (solid lines)  and for time-independent parabolic $ V_{\rm n} $  (dashed lines)  for $T=1.6$K and $1.9$K.(Color Online) }
\end{figure*}

\paragraph{Implementation Details}
The simulations were set up in a planar channel of the size $4h\times 2h\times2h, h=0.05$ cm. The vortex-lines dynamics 
was solved using VFM\cite{Schwarz88,b:Kond-1,Dima15} with 4th-order difference scheme for the derivatives $\B s'$ and  $\B s''$ \cite{BB2011,5point}. We used the periodic boundary condition in the streamwise($x$) and  the spanwise($z$) directions, with slip conditions in the wall-normal $y$ direction, the line-resolution $\Delta \xi_0=0.0016$ cm and the timestep $\delta t\sb s$, defined by the stability condition of the 4th-order Runge-Kutta scheme.
The dynamics of the mean normal velocity profile $V\sb n (y,t)$, \Eq{VnEq}, was solved using 2nd-order finite-difference scheme for the viscous term, 2nd-order Adams-Bashforth method for the time marching and no-slip conditions on the solid walls. The effective pressure gradient $P\sb x=dP/dx$ was used as a free parameter of the system.

To calculate the mutual friction term, we used  two-stage approach. The term $\C F\sb{ns}$ in \Eq{VnEq} implies averaging over space and time, while the term $ F\sb{ns}$ is instantaneous and is averaged only over space.
The parabolic normal velocity profile was initiated on a grid of a mesh size $\delta y=0.00625$ cm in the wall-normal direction. Using this profile, the superfluid system was propagated for $\Delta t=10 \delta t\sb s$. At each time step, the mutual friction force $ F\sb{ns}$ was integrated over thin slices of volume $4h\times\delta y\times 2h$ and  then averaged during  $\Delta t$. Resulting values of $\C  F\sb{ns}$ were assigned to the middle points of the slices in $y$-direction and linearly interpolated to the normal velocity grid points. With this mean mutual friction force, $V\sb n(y)$ was then propagated to $\Delta t$ using the same $\delta t\sb s$. The newly obtained values of $V\sb n(y)$ were interpolated to the positions of the line points for the next cycle of the superfluid dynamics, during which $V\sb n$ was considered constant.

The resulting profiles of $\C L(y)$, $V\sb {s,n}(y)$ and other related properties were obtained by integrating over the same slices as the mutual friction, assigned to the middle points of the slices and then averaged over more than 50 steady-state configurations.
The simulations were carried out for $T=1.45, 1.6$ and $1.9$ K and a number of $P\sb x$ values. The material properties of the fluid components \cite{ExpData} are given in the Table \ref{TabPar}. Notice that we didn't use the modified mutual friction coefficients\cite{Idowu00}, because we only consider the mean normal velocity field.
\section{Results and Discussion}\label{sec:res}
The steady-state profiles of $V\sb n(y)$ for different $T$ and values of effective pressure are shown in  Fig.~\ref{fig:Vnprofiles}. To compare the shapes of profiles we normalized them by  $3/2~\langle V\sb n\rangle$. For the original parabolic profile this value corresponds to the centerline velocity. For modified profiles this relation no longer valid. The profiles of $V\sb n(y)$ become flatter in the center of the channel, increasingly so with increasing temperature and the applied pressure gradient. The resulting profile of counterflow velocity $V\sb{ns}(y)$ is shown by solid lines in Fig.~\ref{fig:Lprofiles}, left panel, for $T=1.6$K and $1.9$K together with profiles, obtained using time-independent parabolic profile with  similar $\langle V\sb {n}\rangle$(dashed lines). The flatter profiles of $V\sb n$ lead to a change in the profiles of $V\sb {ns}$,  especially in the channel core. The vortex lines tend to concentrate in the regions with smaller  $V\sb n$\cite{Dima15}, such that the peaks in the VLD profiles are pushed further toward the walls.

The statistical properties of the vortex tangle are commonly characterized by relation between  mean VLD in the channel and $\langle V\sb{ns}\rangle$:
$\langle \mathcal{ L} \rangle^{1/2}=\gamma(T) (\langle V_{ns}\rangle-v_0)$, where $\gamma(T)$ is a temperature-dependent coefficient and $v_0$ is  virtual origin. This relation is valid only globally\cite{MechMom16}.  We compare in Table \ref{tab:1} the values of $\gamma$, obtained with different profiles. Notice that results for different profiles at temperatures $T=1.45$K and $1.6$K are close, while for higher $T=1.9$K, $\gamma\sb c$ is significantly larger than $\gamma\sb p$. The values of $\gamma\sb p$ are close to those of uniform $V\sb n$ for all temperatures. Notably, our results are higher than those, obtained in Refs.\cite{BL15,YT15} for the channel counterflow with parabolic and Hagen-Poiseuille profiles. This discrepancy may stem from the fact that in both \cite{BL15,YT15} the counterflow condition did not include contribution from the vortex tangle, which is not negligible for dense tangles.

\paragraph{Model profile of $V\sb n$}

To rationalize the observed modifications of the normal velocity profile, we notice that in steady state the mutual friction force is almost constant across the channel, except for the near-wall region, where it quickly falls to zero. Therefore, qualitatively, the mutual friction redefines the  effective pressure gradient in the middle of the channel $P_x+\langle \C{F}\sb{ns} \rangle/\rho\sb n$, while near the wall it remains $P_x$. Such a change leads to flattening of the normal velocity profile (as compared to classic parabolic profile).

To find new, flattened, $ V\sb n$ we first model the shape of mutual friction force profile by a function:
 \begin{equation}\label{Fmodel}
{\C F}\sb{ns}=-(1-{y^\dagger}^n)  \C F\sb{ns}(0) \, ,
 \end{equation}
 where $n$ is an even integer and  $\C F\sb{ns}(0)$ is the mutual friction force in the middle of the channel. To account for the sharp transition from  almost a constant value in the middle of the channel to zero at the wall, large values of $n$ are required. In this work we use $n=14$.
 Substituting \eqref{Fmodel} into steady state equation \eqref{VnEq} for $V_n$  we get:
\begin{equation}\label{Vn2}
\frac{\nu}{h^2} \frac{d^2 V\sb n}{d{y^\dagger}^2}=-P\sb x+\frac{ \C{F}\sb{ns}(0)}{\rho_n}(1-{y^\dagger}^n)\, .
\end{equation}		
This equation is easily solved, giving:
 \begin{eqnarray}\label{Vn3}
 &&V\sb n(y^\dagger)=A\left(1-{y^\dagger}^2\right)\left[1+\frac{2C}{(n+2)(n+1)}\frac{1-{y^\dagger}^{n+2}}{1-{y^\dagger}^{2}}\right]\, ,\\\nonumber
 &&A=h^2P\sb{x,m}/(2\nu) \, , C= \C{F}\sb {ns}(0) /(\rho_n P\sb{x,m})\, ,P\sb{x,m}=P\sb x- \C{F}\sb{ns} (0)/\rho_n\, .
\end{eqnarray}
Notice, that when mutual force is not taken into account $ \C{F}\sb{ns} (0)=0$, the profile \eqref{Vn3} reduces to the classical parabolic profile. To proceed, we notice that $\langle V_n \rangle =2A/3  +2AC/[(n+1)(n+3)]$ and $\langle \C{F}\sb{ns} \rangle=n\C{F}\sb{ns} (0)/(n+1)$. Using counterflow condition and Gorter-Mellink  relation between mean mutual friction force and $\langle V\sb{ns} \rangle$, we find
\begin{equation}\label{GM}
\langle \C{F}\sb{ns} \rangle =A\sb{GM} \rho\sb s \rho\sb n (1+\rho\sb n/\rho\sb s)^3 \langle V\sb n \rangle^3\equiv\frac{\rho_n \nu}{h^2}\frac{\langle V_n \rangle^3}{V_0^2}\, .
\end{equation}
The characteristic velocity $\displaystyle V_0=\left(A\sb{GM} \rho_s h^2(1+\rho\sb n/\rho\sb s)^3 /\nu \right)^{-1/2}$ corresponds to a balance between mutual friction and viscous terms and $A\sb{GM}$ is the Gorter-Mellink constant.
Combining all above we finally  get:
\begin{eqnarray}\label{Vn4}
V\sb n(y^\dagger)&=&\frac{3}{2}\langle V\sb n \rangle \left[V_1 (1-{y^\dagger}^2)+V_2\left(1-{y^\dagger}^{n+2}\right) \right]\, ,
\end{eqnarray}
where $V_1=\Big(1-\frac{1}{n(n+3)}\Big[\frac{\langle V\sb n\rangle}{V_0}\Big]^2\Big)$ and $V_2=\frac{2}{3(n+2)n}\Big[\frac{\langle V\sb n\rangle}{V_0}\Big]^2$ are $T$-dependent expansion coefficients.
The  model profiles are compared with numerical results in Fig\ref{fig:Vnprofiles}, left panel.
\begin{table}[t]
\caption{The values of $\gamma(T)$ for different profiles: $\gamma\sb{c}$ for coupled dynamics, $\gamma\sb{p}$ for frozen parabolic profile. For comparison we give also values for the parabolic profile, Ref.\cite{BL15}, the  Hagen-Poiseuille profile $\gamma\sb{hp}$\cite{YT15} and for the uniform normal velocity $\gamma\sb{uni}$\cite{b:Kond-1}.}
\label{tab:1}       
\begin{tabular}{|c|c|c|c|c|c|c|c|}
\hline\noalign{\smallskip}
T(K) & $\gamma\sb{c}$ (s/cm$^2$)&  $\gamma\sb{p}$(s/cm$^2$) &$\gamma\sb{p}$(s/cm$^2$) \cite{BL15}&$\gamma\sb{hp}$(s/cm$^2$)\cite{YT15}&$\gamma\sb{uni}$(s/cm$^2$)\cite{b:Kond-1} \\
\noalign{\smallskip}\hline\noalign{\smallskip}
1.3 & - & - &$67.9$&$31$&$72.1$\\
1.45 & 83 & 85 &$-$&$-$&$-$\\
1.6 & 114 & 113&$83.6$&$47$&$115.7$ \\
1.9 & 165 & 144 &$105.7$&$103$&$148$ \\
\noalign{\smallskip}\hline
\end{tabular}
\end{table}
\section{Conclusions}
We have studied back-reaction of the quantum vortex tangle on the mean  laminar normal velocity profile in the channel counterflow. The initially parabolic $ V_n $ evolves to a flatter shape. The  vortex tangle influences the mean normal velocity profile via mutual friction force. As a result, both the global properties of flow and the profiles of microscopic properties of the tangle, such as rms curvature, change compared to the uncoupled dynamics, with effect being stronger for high temperatures and counterflow velocities. At low and moderate temperatures the flow and tangle properties are close to those with the time-independent normal velocity profile. We propose a model of $ V\sb n(y)$, expressed via $\langle V\sb n\rangle$, that accounts for the effect of mutual friction.
\begin{acknowledgements}
We acknowledge useful and encouraging discussions with V.L'vov and I. Procaccia.
\end{acknowledgements}



\end{document}